\documentstyle[preprint,aps]{revtex}
\begin{document}

\draft
\tighten

\preprint{\vbox{
\hbox{CWRU-P19-1996}
}}

\title{Family Replication in the Dual Standard Model}

\author{Hong Liu\footnote{hxl20@po.cwru.edu},
Glenn D. Starkman\footnote{gds6@po.cwru.edu} and 
Tanmay Vachaspati\footnote{txv7@po.cwru.edu}}
\address{
Physics Department\\
Case Western Reserve University\\
Cleveland OH 44106-7079.
}

\date{\today}

\maketitle

\begin{abstract}

The family replication problem is addressed in the context
of the dual standard model.  The breaking of
a simple grand unified group to $[G_{low} \times H_1 
\times H_2 \times H_3]/Z_5^3$ and then further to  $G_{low}$, 
produces a spectrum of stable monopoles that fall in three
families each of whose magnetic quantum numbers correspond 
to the electric charges on the fermions of the Standard Model. 
Here $G_{low}=[SU(3) \times SU(2) \times U(1)]/Z_6$ is the 
symmetry group of the standard model above the weak scale, 
and $H_i$ are simple Lie groups which each have a  $Z_5$ symmetry 
in common with $G_{low}$. 

\end{abstract}




In his 1962 paper, Skyrme \cite{skyrme} made the radical 
suggestion that it may be possible to develop a model of baryons 
and mesons as the solitons of what is now known as 
the Skyrme model. More recently, it has been realized 
\cite{tv,hltv} that one can take Skyrme's program yet further 
and construct a ``dual standard model'' in which the magnetic 
monopoles of the model correspond to the quarks and 
leptons of the standard model. 
In this program, there is no freedom to add particles to the 
spectrum since the spectrum 
of monopoles is completely determined by the topology of the 
model. If successful, such a
program would correctly reproduce the charge 
spectrum of standard model fermions, the group representations 
in which they fall, the correct spacetime transformation
properties, and, ultimately, the mass spectrum
and dynamics of interacting fermions, thus making it possible
to answer questions that current particle physics models do
not even attempt to address.  

In \cite{tv,hltv} it was pointed out that breaking a Grand Unified
$SU(5)$ symmetry results in a charge spectrum
of stable magnetic monopoles that is in one-to-one correspondence
with a single family of standard-model fermions. 
The symmetry breaking under consideration was:
\begin{equation}
SU(5) \longrightarrow
 [SU(3)\times SU(2)\times U_Y(1)]/Z_6 \ 
\label{su5break}
\end{equation}
and the scalar field masses were chosen so that the long range
$SU(3)\times SU(2)$ interactions between monopoles is stronger 
than the $U_Y(1)$ interactions.
Then the charge spectrum of the stable monopoles in this model 
is shown in Table 1 together with the charge spectrum of the
first family of standard model fermions.

\

\noindent TABLE I.
\small
Charges on classically stable $SU(5)$ monopoles and
on standard model fermions. The monopole
degeneracies $d_m$ and the number of fermions with a
given set of charges, $d_f$, are also shown. 
\normalsize
\begin{center}
\begin{tabular*}{16.0cm}{|c@{\extracolsep{\fill}}cccc|ccccc|}
\hline
&&&&&&&&&\\[-0.15cm]
                  {$n_{~}$}
                 & {$n_3$}
                 & {$n_2$}
                 & {$n_1$}
                 & {$d_m$}
                 & {$$}
                 & {$SU(3)_c$}
                 & {$SU(2)_L$}
                 & {$U(1)_Y$}
                 & {$d_f$~}             \\[0.20cm]
\tableline
&&&&&&&&&\\[-0.1cm]
+1&1/3&1/2&+1/6&6&$(u,d)_L   $ &1/3 &1/2  &+1/6&6        \\[0.35cm]
-2&1/3&0  &-1/3&3&$d_R       $ &1/3 &0    &-1/3&3        \\[0.35cm]
-3&0  &1/2&-1/2&2&$(\nu ,e)_L$ &0   &1/2  &-1/2&2        \\[0.35cm]
+4&1/3&0  &+2/3&3&$u_R       $ &1/3 &0    &+2/3&3        \\[0.35cm]
-6&0  &0  &-1  &1&$e_R       $ &0   &0    &-1  &1        \\[0.25cm]
\tableline
\end{tabular*}
\end{center}
\vspace{0.3cm}

While the correspondence between the monopoles of the $SU(5)$
model and a family of the standard model fermions is quite remarkable, 
it is by no means complete since we know that the standard model has 
three families of fermions and not just one. The existence of three 
families of light fermions has long been an outstanding problem in 
particle physics \cite{FamUni}. It is this problem that we now 
address: we find a symmetry breaking that yields three families of 
monopoles with the magnetic charge spectrum of each of these families 
corresponding exactly to the electric charge spectrum of a family of 
standard model fermions.

The strategy we adopt for obtaining three families of monopoles
is to build upon the correspondence of $SU(5)$ monopoles
with quarks and leptons. In essence, we want the $SU(5)$ monopoles
three times over. For this, we consider a symmetry breaking
pattern of the kind:
\begin{equation}
G \rightarrow K \equiv \biggl [ 
 SU_0(5)\times H_1 \times H_2 \times H_3 \biggr ] / 
   [ Z_5^{(1)} \times Z_5^{(2)} \times Z_5^{(3)} ]
\label{pattern}
\end{equation}
where, $G$, $H_1$, $H_2$ and $H_3$ are all simply connected groups.
(The $Z_5$ factors in the denominator can be generalized to $Z_n$
where $n \ge 5$ but $n\ne 6$. Here we will only consider the choice 
$n=5$ as it is the simplest.)
The $Z_5^{(i)}$ ($i=1,2,3$) contain group elements that are common
to $SU_0(5)$ and $H_i$. A specific example of such a symmetry breaking
pattern is $H_i = SU_i(5)$, with $Z_5^{(i)}$ the 
center of $H_i$. Then $K = SU(5)^4/Z_5^3$, that is, all four 
$SU(5)$'s share a common $Z_5$ center. 
A possible choice for $G$ is $SU(5^4)=SU(625)$ but smaller groups may
also work \footnote{We thank John Preskill for giving us this
example of $G$.}. Since the spectrum of monopoles 
depends only on the incontractable closed paths in $K$, 
the actual choice of
$G$ is immaterial so long as $G$ is simply connected. 

Consider the incontractable paths in $K$. An example of such
a path is one that starts on the identity, traverses 
$SU_0(5)$ to an element of $Z_5^{(i)}$, then returns to the identity
through $H_i$.  This is a closed path
that is incontractable because of the discrete nature of $Z_5^{(i)}$
and corresponds to a monopole with $SU_0(5)$ and $H_i$ charge.
We call this monopole a ``digit.'' Similarly,
there are paths that pass through $H_i$ and $H_j$ ($i\neq j$)
and avoid $SU_0(5)$ altogether; these correspond to a monopole which
is a singlet of $SU_0(5)$ but which has $H_i$ and $H_j$ charge. 
We refer to these as ``sterile'' monopoles.
All other incontractable paths (and hence all other monopoles), 
such as those that pass through $SU_0(5)$ and several of the $H_i$,
can be built out of these two types of paths. 

We next break the $SU_0(5)$ to $[SU(3)\times SU(2)\times U_Y(1)]/Z_6$,
the symmetry group of the standard model. 
The pre-existing monopoles from the
symmetry breaking in (\ref{pattern}) will now get $SU(3)$,
$SU(2)$ and $U_Y(1)$ charges. In addition, new monopoles lying
entirely in the $SU_0(5)$ sector will be produced since 
$[SU(3)\times SU(2)\times U_Y(1)]/Z_6$ has its own incontractable
closed paths. We will refer to these as ``pure'' monopoles.
The $U_Y(1)$ charge on a pure monopole is 5 times the $U_Y(1)$ charge 
on the digit that has the same $SU(3)$ and $SU(2)$ charge. 
To see this, note that the incontractable path for the digit 
(produced during $G\rightarrow K$) 
need only traverse between elements of $Z_5^{(i)}$ shared 
with the $U_Y(1)$. 
For example, there is a pure monopole corresponding to the path
that traverses the entire $U_Y(1)$ circle. 
But, there is a digit with the same $SU(3) \times SU(2)$ charges
whose path traverses only one-fifth of the $U_Y(1)$ circle 
and then closes by traversing a path in the $H_i$ factors. 
So at this stage there are two types of monopoles:
digits, with non-zero 3-2-1 and $H_i$ charges, 
and, pure monopoles with zero $H_i$ charge 
and a $U_Y(1)$ charge that is 5 times the charge of the digit
with the same $3-2$ charges.

The next step is to break each of the $H_i$ to $Z_5^{(i)}$
since we want the low energy symmetry to be the usual
$[SU(3)\times SU(2)\times U_Y(1)]/Z_6$.
This symmetry breaking does not yield
any new monopoles but it does produce $Z_5$ strings that
confine the digits into clusters of 5 
with each cluster being a singlet of the $H_i$. 
Since this cluster is a singlet of all the $H_i$, 
its topological charge agrees with the topological charge 
of the corresponding pure monopole.
Hence the $SU(3)$ and $SU(2)$ charges, and, the hypercharge on
all the monopoles are given by the usual values shown in Table I. 
At this stage, the sterile monopoles also get connected by
strings into $H_i$ singlets. But since the sterile monopoles
have no $SU_0(5)$ charge, the clusters of sterile monopoles are
topologically trivial and can decay to the vacuum. The exception
to this statement would be if the cluster is fermionic (as
of course we must imagine all the other clusters to ultimately
be if they are to correspond to standard model fermions). 
Fermionic clusters of sterile monopoles would correspond to
right-handed neutrinos.

Now that the charge spectrum of the monopoles agrees with that shown in 
Table 1, we need to count the different monopoles of each 3-2-1
charge. For this we look at the interactions of the digits
in a cluster. The digits interact by exchange of 3-2-1 gauge and
scalar fields and, by an argument identical to that in \cite{gh}, 
a cluster of 5 digits would be unstable to de-clustering in the
absence of $Z_5$ strings. But the $Z_5$ strings provide a confining 
potential and do not allow the cluster to disperse. This shows that
the pure monopoles are unstable to decaying into a cluster of digits
that are confined by strings. 

The digit clusters confined by $Z_5$ strings in each of the 
three $H_i$'s will turn out to be the three families of monopoles
that correspond to the three families of standard model fermions.
The clusters composed of digits having charges in
different $H_i$'s are unstable to decay into a cluster with
digits having charge in a single $H_i$. 
We show this by explicit construction in a 
concrete example.

First we realize that nothing changes if $SU_0(5)$ is replaced
by $[ SU(3)\times SU(2) \times U_Y(1) ]/Z_6$ directly in 
eq. (\ref{pattern}) since the $Z_5$ center of $SU_0(5)$ is contained 
in $U_Y(1)$.  Let us now consider the specific symmetry breaking
\begin{eqnarray}
G & \rightarrow \biggl [ [ SU(3)\times SU(2) \times U_Y(1) ]/Z_6
                  \times SU(5)^3 \biggr ] /Z_5^3 \\
\label{specific0}
& \rightarrow [ SU(3)\times SU(2) \times U_Y(1) ]/Z_6 \ .
\label{specific}
\end{eqnarray}
The monopoles formed in the first stage of symmetry breaking
correspond to all closed incontractable paths in the unbroken
group which have the form
$$
P[s] = {\rm exp} \biggl [ i s (n_3 T_8 + n_2 \lambda_3 + n_1 Y 
                     + \sum_{i=1}^3 m_i \Lambda^i_{24}
          \biggr ]\ , \ \ \ s\in [0, 4\pi ]
$$
where, $n_i$ and $m_i$ are integers. The generators
$T_8$, $\lambda_3$, $Y$ and $\Lambda_{24}^i$ of $SU(3)$, $SU(2)$,
$U_Y(1)$ and $SU_i(5)$ ($i=1,2,3$) respectively generate
the centers of these groups. They are normalized to satisfy 
$$
e^{ i 4\pi n T_8 } = e^{- i2\pi n /3} {\bf 1}
$$
$$
e^{ i 4\pi n \lambda_3 } = e^{i2\pi n /2} {\bf 1}
$$
$$
e^{ i 4\pi n Y } = e^{i2\pi n /30} {\bf 1}
$$
$$
e^{ i 4\pi n \Lambda_{24}^j } = e^{i2\pi n /5} {\bf 1} \ ,
\ \ \ {j=1,2,3}
$$
where, $n$ is any integer and ${\bf 1}$ is the identity element 
of $G$. For $P[s]$ to be closed we need $P[0] = P[4\pi ]$ and
so we have the following constraint on the integers $n_i$, $m_i$:
\begin{equation}
 {{n_1} \over 30} + {{n_2} \over 2} - {{n_3} \over 3} 
+ {{m} \over 5} = {\rm integer}
\label{condition1}
\end{equation} 
where $m = m_1 + m_2 + m_3$.
The only monopoles in which we are interested are those with 
non-trivial hypercharge ($n_1$) since those with $n_1 =0$ will 
be topologically
equivalent to the vacuum once the $SU(5)$'s break down in the
second stage of symmetry breaking. So $n_1 \ne 0$.
Now we want to find all possible $n_i$, $m_i$ so as to satisfy
(\ref{condition1}) with $n_1 \ne 0$. 

We would like to restrict our attention to only those solutions 
that lead to stable monopoles. Following \cite{gh,tv,hltv}, we
will consider scalar field masses such that the long range
$SU(3)\times SU(2)$ interactions are much stronger than the 
$U_Y(1)$ interactions. We will further assume mass parameters
such that the $U_Y(1)$ interactions are much stronger than the 
$SU(5)^3$ interactions. With this assumption, the $SU(5)^3$
interactions do not play any role in the stability analysis
of monopoles with $n_1 \ne 0$ and the results in \cite{gh}
apply directly. Hence the monopoles with $n_1 = 5$ and 
$n_1 > 6$ are unstable to decay. (Similiarly for negative $n_1$.)
This means that we can restrict our attention to $n_1 = 1,...,6$. 

Note that if we do find a solution, adding 3 to $n_3$, 2 to $n_2$,  
30 to $n_1$, or 5 to $m$ will also yield a solution.
These additional solutions correspond to adding closed paths
that are trivial in the case of $n_3$ and $n_2$ and non-trivial
in the case of $n_1$. In the case of adding 5 to $m$, the additional 
closed path may be trivial or non-trivial depending on how the 5 is
split between the $m_i$. But, since the $SU(5)$'s will ultimately
be broken, the monopoles corresponding to the non-trivial closed
paths in the case of $m$ will cluster in topologically trivial
configurations. So we restrict our attention to:
$$
n_3 = 0,\pm1 \ , \ \ \ n_2 = 0,1 \ , \ \ \ n_1 = 1,2,...,6 \ ,
\ \ \ {\rm and} \ \ \ m=0,\pm1,\pm2
$$
Then the monopole solutions from the first stage of symmetry 
breaking are shown in Table II. We have also indicated the
stability of each monopole.

\begin{center}
TABLE II.
\small{
Solutions for $n_i$, $m$ for the digits and stability of 
the digit.
}
\end{center}
\normalsize
\begin{center}
\begin{tabular*}{6.0cm}{|c@{\extracolsep{\fill}}ccc|c|}
\hline
                   \multicolumn{1}{|c}{$n_{1}$}
                 & \multicolumn{1}{c}{$n_2$}
                 & \multicolumn{1}{c}{$n_3$}
                 & \multicolumn{1}{c|}{$m$} 
                 & \multicolumn{1}{c|}{stable?} \\
\hline
1&1&1&-1&yes \\
2&0&-1&-2&yes \\
3&1&0&2&yes  \\
4&0&1&1&yes  \\
5&1&-1&0&no   \\
6&0&0&-1&yes \\
\hline
\end{tabular*}
\end{center}
\vspace{0.3cm}

Consider first the masses of the digits shown in Table II.
The $n_1=1$ digit has $m=-1$ and so could be 
any one of the three forms:
$(m_1,m_2,m_3)=(-1,0,0)$, $(-1,-1,1)$, $(-1,-2,2)$
(or permutations thereof). If we assume that the mass of
a monopole is proportional to the charge as in the BPS case,
this tells us that the square of the masses go like:
\begin{equation}
\begin{array}{rl}
M_1^2 &= Tr((-\Lambda^{(1)}_a)^2)\\
M_2^2 &= Tr((-\Lambda^{(1)}_a)^2) 
     + Tr((-\Lambda^{(2)}_{a'})^2) 
     + Tr((\Lambda^{(3)}_{a''})^2)\\
    &= 3 M_1^2\\
M_3^2 &= Tr((-\Lambda^{(1)}_a)^2) 
     + Tr((-\Lambda^{(2)}_{a'}-\Lambda^{(2)}_{b'})^2) 
     + Tr((\Lambda^{(3)}_{a''}+\Lambda^{(3)}_{b''})^2) \\
    &= 5 M_1^2 + 4Tr(\Lambda^{(i)}_a\Lambda^{(i)}_b) \\
\end{array}
\end{equation}
While $Tr(\Lambda^{(i)}_a\Lambda^{(i)}_b) < 0$, ($a\ne b$),
$\vert Tr(\Lambda^{(i)}_a\Lambda^{(i)}_b)\vert  \le 
Tr((\Lambda^{(i)}_a)^2) = M_1^2$
(in SU(5),  this inequality is $5 < 20$), so $M_2,M_3 > M_1$.  
Thus the lightest $n_1=1$ digit is indeed 
$(m_1,m_2,m_3)=(-1,0,0)$ or $(0,-1,0)$, $(0,0,-1)$.
An equivalent calculation shows that 
the lightest  $n_1=2$ digit (up to the three permutations) is 
$(m_1,m_2,m_3)=(-2,0,0)$; the lightest $n_1=3$ digit is 
$(m_1,m_2,m_3)=(2,0,0)$; the lightest $n_1=4$ digit is 
$(m_1,m_2,m_3)=(1,0,0)$; the lightest $n_1=6$ digit is 
$(m_1,m_2,m_3)=(-1,0,0)$.
Therefore the lightest digits come in three families
with the family identified by the $i$ for which $m_i$
is non-zero.

In addition to the digits shown in Table 2,  
there are sterile monopoles for which $n_i =0$ but $m_i \ne 0$. 
These will form topologically trivial 
clusters once the $SU(5)$'s break. 
There are also the pure monopoles for which 
$m_i =0$ but $n_i \ne 0$. For these pure monopoles, we can
only have $n_1 =5, 10, 15, 20, 25, 30$. We have already
argued (below eq. (\ref{condition1})) 
that all of these are unstable because the $n_1=5$ monopole
is unstable and the others have $n_1 > 6$.
The $n_1=5$ monopole is unstable to fragmentation into
an $n_1=2$ and a $n_1=3$ monopole since these two monopoles
mainly interact by the repulsive hypercharge interaction. 
For a similar reason, the $n_1=n_* >6$ are unstable to 
fragmentation into an $n_1=6$ and an $n_1=n_*-6$ monopole.
The instability of charge 5 monopoles in our model is crucial 
to the realization that the model contains three families of monopoles.

When the $SU_i(5)$ break to $Z_5$, the digits must bind
into clusters with trivial $SU_i(5)$ charge, i.e. $SU_i(5)$ charge  
which is a multiple of 5. In the cluster of $|m| =1$ digits,
the $m$'s on each digit will all live in the same $SU_i(5)$
since they have to be confined by $Z_5$ strings belonging
to the same $SU_i(5)$ factor. In the cluster of 
$\vert m\vert = 2$ digits, we could have five $m=2$ digits, 
or, five $m_i=1$ and five $m_j=1$ digits. Since an $m=2$ digit 
is lighter than
two $m=1$ digits (the $m=2$ digit is stable to fragmentation
into two $m=1$ digits) we expect the five $m=2$ digits to be
less massive than the ten $m=1$ digits. 
These five digits must have charges in the same
$SU_i(5)$ to be confined by $Z_5$ strings. In general, a cluster 
of digits other than the lightest digits will decay into a
cluster of lightest digits since these are related by differences
of 5 sterile monopoles which are equivalent to the vacuum.

We would next like to determine the $SU(2)$ and
$SU(3)$ arrangements of the clusters. 

Consider first the $SU(2)$ arrangement of a cluster of five 
$n_1=1$ digits.  These could take any one of the five forms:
\begin{eqnarray*}
&(UUUUU)_i =5U_i\ ,\ (UUUUD)_i=4U_i+D_i\ ,\ (UUUDD)_i=3U_i+2D_i\ ,\\ 
&(UUDDD)_i=2U_i+3D_i\ ,\ (UDDDD)_i=U_i+4D_i\ ,\ (DDDDD)_i=5D_i   
\end{eqnarray*}
where $U_i$ is the $n_2=+1$, $m_i \ne 0$ and $D_i$ is the $n_2=-1$,
$m_i\ne 0$ digit. (We have suppressed the $SU(3)$ labels for
convenience.)
However, while there is an attractive $SU(3)\times SU(2)$
force between both two
U's and between a U and a D, the latter is stronger \cite{hltv}, 
and hence the lowest energy configurations will be
\begin{equation}
3U_i+2D_i \ , \ \ \ {\rm and}, \ \ \ 2U_i+3D_i \ .
\end{equation}
The former we identify as being dual to the 
$u_L$, $c_L$ and $t_L$ quarks (for $i=1,2,3$), the latter to the 
$d_L$, $s_L$ and $b_L$ quarks. (In what follows, we shall suppress
the index $i$ which tells us the variety of $Z_5$ strings that
connect the digits within the cluster.)

Similarly we could consider the $SU(3)$ arrangement of the 
$3U+2D$ cluster. If we label the $SU(3)$ charges by $b$, $g$
and $r$, the most tightly bound cluster will have color 
arrangements of the kind $2b+2g+r={\bar r}$, or, $2b+g+2r={\bar g}$,
or, $b+2g+2r={\bar b}$. These arguments are straightforward to 
apply to all the other clusters discussed below and we shall not 
state them explicitly. 

The $n_1=2$ digit, is an $SU(2)$ singlet, and can
be thought of as ${\bar D}_L=U+D$.
The cluster of $n_1=2$ digits 
which we identify as the ${\bar d}_L$, ${\bar s}_L$ and 
${\bar b}_L$ anti-quarks can be thought of as
$5{\bar D}_L = 5U+5D$.

The $n_1=3$ digit is again an $SU(2)$ doublet and so comes
in two forms ${\bar E}_R=2U+D$ and ${\bar N}_R=U+2D$ 
which we identify as respectively the right-handed anti-electron
digit and the right-handed anti-neutrino digit.
The cluster of $n_1=3$ digits once again has 2 possible forms
$3{\bar E}_R+2{\bar N}_R =8U+7D$, and, 
$2{\bar E}_R+3{\bar N}_R =7U+8D$,
which we identify as being
dual to the ${\bar e}_R$, ${\bar \mu}_R$${\bar \tau}_R$
and to ${\bar \nu}_{eR}$,  ${\bar \nu}_{\mu R}$, and,
${\bar \nu}_{\tau R}$ respectively.

The $n_1=4$ digits, $U_R=2U+2D$,
which we identify as the right-handed up-quark digit 
clusters according to $5U_R =10U+10D$. This is dual to 
$u_R$, $c_R$, and, $t_R$.

A cluster of 5 $n_1=5$ digits is unstable to decay.
The $n_1=6$ digit, $E^+_L = 3U+3D$
clusters to form $5E^+_L = 15U+15D$ which is dual to
$e^+_L$ (left-handed positron), $\mu^+_L$, and, $\tau^+_L$.

We have now identified the fundamental fermions of the standard model
and demonstrated how the triplication occurs dynamically.
One might notice that there is no clear prediction of the existence
or absence of right-handed neutrinos, since these are topologically
trivial (at least in the 3-2-1 sector). However, there are certainly
many potential candidates, namely the clusters of sterile monopoles.

In \cite{tv,hltv} several issues not resolved in the earlier
(or present) version of the dual standard model were pointed out.
These issues had to do with the spacetime transformation properties
of monopoles - that is, the spin and chirality of monopoles. It is 
conceivable
that a resolution of these problems will indicate that the spectrum
of monopoles we have found will have additional degeneracies. For
example, there could be monopoles with the same internal charges
but with different spins. In fact, such a degeneracy might account
for the electroweak Higgs since it has the same internal charges
as the electron-neutrino doublet. The issue then would be to 
investigate why the monopole field dual to the electroweak Higgs 
acquires a vacuum expectation value. These issues are hard to
address since they are non-perturbative but we hope that
they can be addressed within a lattice formulation and studied
analytically in a supersymmetric context.

Some phenomenological issues arise in the dual standard
model that we now address. The first issue is that the monopole 
corresponding to the proton has the same charges as the monopole 
corresponding to the positron and so topology
does not forbid this transition: proton decay should be 
possible. But baryon and lepton numbers are approximately
conserved in the standard model and so the proton decay
rate in the dual standard model had better be suppressed. 
The suppression of the decay rate can only come from
dynamical arguments. This is possible at the classical level if
there is an energy barrier that prevents three loosely clustered
monopoles that we would identify with a proton from collapsing and 
forming a tightly bound monopole that we would identify with a positron. 
Such a barrier is indeed present as can be seen
by constructing the interaction potential between
an $n=1$ and an $n=2$ $SU(5)$ monopole as done in \cite{gh,hltv}. 
It would be of interest to see if this barrier survives on going 
beyond the classical level calculation. Another issue
is that of the rate of flavor changing processes. In the present
model, for example, a $t$ quark monopole can indeed convert to an 
$u$ quark, but the process requires an intermediate state that is
a pure monopole with the charges of an $u$ quark. We know that the
pure monopole has higher energy than both the $t$ and $u$ quarks.
So the decay of the $t$ monopole to the $u$ monopole is a classically
forbidden but quantum mechanically allowed process. 

Within the philosophy of the dual standard model, it is interesting
to note that $SU(5)$ cannot be the ultimate symmetry of particle 
physics since it does not yield the three families of particles
that we know to exist. If the model described in this paper is
the only way to get three families, it tells us that the true symmetry
group must be large enough to contain
$$
\biggl [ [ SU(3)\times SU(2) \times U_Y(1) ]/Z_6 \times  
         H_1 \times H_2 \times H_3 \biggr ] /
   [ Z_5^{(1)} \times Z_5^{(2)} \times Z_5^{(3)} ] \ .
$$
Another important prediction of the current model is that the digits,
and not the quarks and leptons, are the fundamental building blocks
of matter. Ultimately we should see these preonic components in the
laboratory.

The successful resolution of the family replication problem in the
dual standard model offers a glimmer of hope that the spectrum
of standard model fermions can indeed be understood in terms of the
topology of certain manifolds.
To us it seems that this is not unlike 
the classification of baryons and mesons in terms of group
representations \cite{gmn}. It is too early to say, however, if the
present attempt will meet with the same degree of success.

{\it Acknowledgements:} GDS is supported by an NSF CAREER award
and TV by the DOE.

\end{document}